# Single Flux Quantum Circuit Operation at Millikelvin Temperatures

Jason Walter, Adam C. Weis, Kan-Ting Tsai, Meng-Ju Yu, Naveen Katam, Alex F. Kirichenko, Oleg A. Mukhanov, Shu-Jen Han and Igor V. Vernik

*Abstract*—**As quantum computing processors increase in size, there is growing interest in developing cryogenic electronics to overcome significant challenges to system scaling. Single flux-quantum (SFQ) circuits offer a promising alternative to remote, bulky, and power-hungry room temperature electronics. To meet the need for digital qubit control, readout, and co-processing, SFQ circuits must be adapted to operate at millikelvin temperatures near quantum processors. SEEQC's SFQuClass digital quantum management approach proximally places energy-efficient SFQ (ERSFQ) circuits and qubits in a multi-chip module. This enables extremely low power dissipation, compatible with a typical dilution cryostat's limited cooling power, while maintaining high processing speed and low error rates. We report on systematic testing from 4 K to 10 mK of a comprehensive set of ERSFQ cells, as well as more complex circuits such as programmable counters and demultiplexers used in digital qubit control. We compare the operating margins and error rates of these circuits and find that, at millikelvin, bias margins decrease and the center of the margins (i.e., the optimal bias current value) increases by ~15%, compared to 4.2 K. The margins can be restored by thermal annealing by reducing Josephson junction (JJ) critical current $I_c$. To provide guidance for how circuit parameters vary from 4.2 K to millikelvin, relevant analog process control monitors (PCMs) were tested in the temperature range of interest. The measured JJ critical current (of the PCM JJ arrays) increases by ~15% when decreasing temperature from 4.2 K to millikelvin, in good agreement with both theory and the empirically measured change in the center of bias margins for the tested digital circuits.**

## I. INTRODUCTION

IN the quest to build fault-tolerant quantum computing (FTQC) that executes useful quantum algorithms, millions of physical qubits need to be integrated [1,2]. Since each qubit requires multiple control lines, this leads to fundamental scaling problems in both physical system integration and data communication. First, connecting cryogenic qubits to their room temperature control and readout electronics via coaxial cables leads to unfeasible cable/connector densities and heat loads [3]. The associated signal distortions, noise, crosstalk, and long latencies need to be mitigated. Second, the active quantum error correction (QEC) protocols for non-Clifford gates would produce syndrome data at rates of terabits/second, which would then need to be transmitted to a room-temperature processor to perform recovery operations at time scales comparable to the coherence time of the qubits. This would require an extremely high-bandwidth, low latency cryo-to-room temperature data link for fast digitization and digital processing [4,5].

These problems can potentially be addressed if fast, low power cryogenic classical control/readout and data processing electronics can be situated in the same cryostat as the quantum processors, obviating the need for coaxial cables, and providing the capability of performing pre-processing to reduce output data bandwidth. The cryoCMOS circuits are being developed to integrate control electronics at a 3 K stage of a dilution refrigerator [6,7]. Still, these electronics are at some nontrivial distance away from the qubit plane at the ~20 mK stage and can only partially address the problems listed above. In contrast, fast and low power superconducting electronics can be integrated much closer to the qubits at the millikelvin stage without overheating [8,9] and could potentially resolve these scaling challenges.

Multiple versions of superconducting circuits are being considered for the implementation of integrated qubit control and readout [9-16]. All are based on superconducting circuit approaches developed for 4.2 K and thus need to be reoptimized for millikelvin operation to minimize power dissipation, circuit size, and potential adverse effects on qubits. The first successful experiments using integrated superconducting circuits at a millikelvin stage demonstrated the validity of this approach [17-20]. However, no systematic study of classical superconducting circuits operating at mK temperatures was performed, leaving many questions unanswered.

This paper reports on the results of comprehensive testing, in the temperature range from 4.2 K to 10 mK, of ERSFQ [21] cells and more complex circuits, including pulse counters and demultiplexers, typically used in digital qubit control circuits.

## II. EXPERIMENTAL APPROACH

We use energy efficient SFQ (ERSFQ) circuits [21] at mK temperatures which are proximally placed in multi-chip modules (MCM) [22] to provide digital control and readout for superconducting qubits. We consistently observe that the performance of circuits designed with the library optimized for a liquid helium (4.2 K) operation significantly changes from 4.2 K to mK temperatures (e.g., optimal bias point and bias margins). To understand the reasons for this change, we designed, fabricated, and tested chips comprised of analog and digital process control monitors (PCMs) and digital ERSFQ circuits of varying complexity across temperature range of interest. Additionally, we studied the ERSFQ demultiplexers and programmable counters used in our experiments of digital qubit control [17]. These chips were fabricated using SEEQC's SFQuClass fabrication process with a critical current density of 10 µA/µm². This is optimized for integration with qubit chips operating in a 20 mK cryogenic environment [23], side-by-side with SEEQC's qubit control and readout chips. This process enables a higher density of energy efficient ERSFQ circuits by lowering the minimum critical current to 10 µA and employing high resistivity junction shunts and a NbN high kinetic inductance layer (HKIL) for bias lines and circuit inductors with a target thickness of 40 nm and measured critical temperature of 11 K. The fabricated chips were first tested in liquid helium at 4.2 K, then in an adiabatic demagnetization refrigerator (ADR) from 4.2 K down to 100 mK, and finally in a dilution refrigerator (DR) at about 10 mK. Importantly, the screening and testing of chips at 4.2 K in liquid helium proceeds much faster than at mK temperatures in ADR and/or DR due to much faster cooldown time and less complex and more developed test infrastructure. Therefore, one of the goals of this study is to be able to credibly predict the performance of ERSFQ circuits at 10 mK solely by testing them at 4.2 K.

### A. Circuit Description

To study ERSFQ circuits at temperatures from 4.2 K to 10 mK and to explain any changes in their performance, we designed a chip with both analog and digital PCMs. Since the ERSFQ circuit performance is mainly influenced by Josephson junction (JJs) critical current and circuit inductances, we decided to study their dependence vs temperature in the range of interest. The analog PCMs include two arrays of ten nominally shunted 20-µA and 100-µA JJs each for monitoring the change of critical current ($I_c$) vs temperature, and SQUID-based inductance test structures [24] to extract characteristic inductances of the NbN and Nb layers and their temperature dependences. The digital PCMs comprise five ERSFQ circuits, ordered by increasing complexity both in terms of JJ count and function: two chains of DC/SFQ-JTL-SFQ/DC with employing DC/SFQ pickup coils; a frequency divider by 4 comprising two serially connected TFFs [25, 26]; a two port D-flip flop [27]; and an NDRO switch comprising an NDRO cell [12, 26] with synchronized set/reset inputs; respectively named sd1, sd2, TFF, D2F and SW in following experimental graphs.

We also tested circuits used in our experiments with digital qubit control at 4.2 K and 10 mK. These include a 1-to-4 demultiplexer (DMX) and an 8-bit programmable counter (PC), both NDRO based. The DMX features a single SFQ digital input that can be routed to four SFQ output channels. Internally, the DMX is a serial-in, parallel-out shift register (SR) with non-destructive readout, with bits stored in the SR activating a path between the input and outputs [29]. Each serial pulse sent to the serial control interface shifts all bits in the SR, while a clock pulse to the input will force the contents of the SR to the outputs, effectively passing the clock pulse to the selected outputs. The PC generates a user-determined number of clock pulses before stopping. Like the fixed counter, the PC [30, 10] has a chain of TFFs to track clock pulses. But whereas the fixed counter always counts to $2^N$, where $N$ is the number of TFFs, the programmable counter can count to any integer up to $2^N$. This is accomplished by programming an NDRO SR with the value $P=2^N-M$, with $M$ the desired count value. Before the count starts, the NDRO SR value is loaded into the TFF chain, which starts from $P$, counting $M$ pulses before activating the stop count output. Together with all input/outputs, the DMX and PC respectively consist of 298 and 617 JJs.

### B. Experimental Setups

Initial prescreening and testing of all structures and circuits was performed in liquid helium at 4.2 K, with chips mounted in a flip-chip immersion cryoprobe using controlled torque to make simultaneous contact to all chip input/output lines.

Both analog and digital PCMs were then tested in ADR at temperatures from 100 mK to 4.2 K with steps of 100 mK, using a thermometer located very near the chips under test. This ADR of base temperature of ~80 mK was manufactured by HPD/FormFactor with a custom-made sample mount closely mimicking that of the flip-chip cryoprobe. The chip temperature was PID controlled by applying current through the ADR magnet with accuracy of ~20 mK at temperatures below 3.5 K and ~100 mK above it, due to more sparse calibration of the thermometer at higher temperatures.

Measurements of more complex ERSFQ circuits, i.e., demultiplexers and counters, were performed in a BlueFors LD400 DR at temperatures of ~10 mK.

All experimental setups (immersion cryoprobe, ADR, and DR) employed multilayer concentric mu-metal cans providing the necessary magnetic shielding.

### III. RESULTS AND DISCUSSION

Figs. 1-3 show measured temperature dependences for analog PCMs. These tests were done in ADR in the temperature range from 100 mK to 4.2 or 4.5 K, and include results obtained in liquid helium at 4.2 K. Three different chips from various locations and wafers were measured. To verify testing reproducibility, one of the chips was measured twice consecutively. To distinguish between wafer-to-wafer and chip-to-chip fabrication variations, all critical current ($I_c$) and inductance results in Figs. 1-3 are normalized to the maximum value for the given array of the given chip. Fig. 1 shows critical current dependences vs temperature $I_c(T)$ for shunted ten-JJ arrays with nominal $I_c$ of 20 µA and 100 µA. These two values of $I_c$ were selected to cover low $I_c$ JJs used in our circuits to control qubits and a higher $I_c$ that is less susceptible to thermal noise at 4.2 K.

For all measured JJ arrays with good reproducibility, Fig. 1 shows monotonic increase of $I_c$ by ~15% when temperature changes from 4.2 K to ~2.5 K, and a subsequent flat response when further cooling down to 100 mK.

Using the Ambegaokar-Baratoff [31] formula for $Ic(T)$:

$$Ic(T) = (\pi \Delta(T)/2eR)\tanh(\Delta(T)/2kT), \qquad (1)$$

the BCS relationship for weak coupling $\Delta(0) \approx 1.76kT_c$ [32], and the BCS-type interpolation formula:

$$\Delta(T) = \Delta(0)\tanh\left(1.74\sqrt{\frac{T_c}{T}-1}\right) \qquad (2),$$

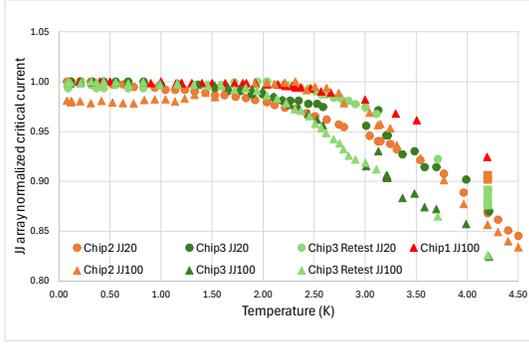

**Fig. 1.** Temperature dependence of normalized critical current for 20 µA and 100 µA arrays of ten JJs. Three different chips with these arrays were tested. See text for details.

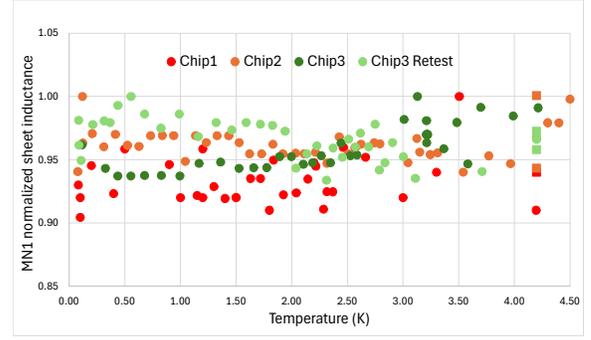

**Fig. 2.** Temperature dependence of normalized sheet inductance for NbN HKIL. Three different chips with these inductances were tested. See text for details.

which approximates the energy gap over our measurement range, one can obtain the following for the dependence of JJ critical current vs temperature $I_c(T)$:

$$\frac{I_c(T)}{I_c(0)} = \frac{\Delta(T)}{\Delta(0)} \tanh\left(\frac{1.76 T_c}{2T} \tanh\left(1.74\sqrt{\frac{T_c}{T}-1}\right)\right) \quad (3),$$

where $R$, $\Delta$, $k$ and $T_c$ are normal state resistance, superconducting gap, Boltzmann constant and critical temperature of JJ, respectively. From (3) and $T_c = 8.5$ K measured for our Nb films, one can get $I_c(4.2K)/I_c(0) \approx 0.88$, which is in good agreement with the $I_c$ change observed in Fig. 1. Note that $T_c$ for the JJs may be further decreased from the continuous-film value, reducing this ratio even more.

Figs. 2 and 3 show temperature dependences for HKIL NbN and other two Nb layers used to fabricate [23] our ERSFQ circuits. Due to pad number limitation on the PCM chip, only one SQUID inductance structure was designed for each the M1 & M2 Nb layers [23] used for small circuit inductors, while multiple structures were designed for HKIL NbN layer MN1 used for large circuit and bias inductors. This allows us to extract a sheet inductance (per square) for HKIL NbN layer in Fig. 2 and only specific inductances (per length) for M1 and M2 Nb layers in Fig. 3 [24]. These SQUIDs have an additional inside lead to allow extraction of 3 inductances from the measured $V(\Phi)$ curves with an approximate 16:9:7 ratio. Note that due to temporary experimental setup malfunction, M1 specific inductance was not measured on chip 2. From the theoretical considerations [33] and measurements [34], the NbN kinetic inductance was not expected to change significantly at $T \ll T_c$. Microwave spectroscopy of NbN resonators performed in [35] in the temperature range of interest revealed a resonance frequency change of less than 0.1% which is caused by the change in inductance. Indeed, the results in Fig. 2 confirm a nearly constant sheet inductance of HKIL NbN film.

Similarly to HKIL layer, Fig. 3 did not show any appreciable temperature variation in the specific inductance for M1 and M2 layers, except for a few outliers. This can be explained by the fact that both M1 and M2 layers are thicker than the Nb London penetration depth [23], and their inductances are mostly determined by geometric inductance.

Figs. 4 and 5 show relevant temperature dependences from 4.2 K to 100 mK measured in ADR for five digital ERSFQ PCMs, with each having a shared bias line for its cell and the input/output monitors. Additional 4.2 K data was also measured in liquid helium and is labelled "He." Fig. 4 displays the normalized center of bias margins (i.e., the optimal bias current value) and Fig. 5 the bias margins themselves. The values for the bias currents in Fig. 4 are normalized to the maximum value of the center of the bias margins measured for that specific PCM. For all ERSFQ PCMs, Fig. 4 exhibits an increase in the normalized center bias by ~15% as temperature decreases from 4.2 K to 100 mK, very similar to that observed in Fig. 1 for $I_c$ of JJ arrays. To highlight the change for the cell bias currents with temperature drop, Fig. 4 shows trendlines for three ERSFQ PCMs: one DC/SFQ-JTL-SFQ/DC, frequency divider by 4, and two port D-flip flop.

Fig. 5 shows the temperature dependences of the full bias current margins of the 5 ERSFQ PCMs. Margins are defined by the range of bias currents for which an error rate of less than 10% was observed in 100 consecutive measurements of the given PCM. This low, non-zero error rate was selected to generate a fair comparison between operations at mK and 4.2 K, accounting for the circuit errors due to thermally activated switching at higher temperatures of JJs with low $I_c$ (on order of 10 µA). The bias margins in Fig. 5 are a combination of two regimes: both the most energy-efficient operation of the circuits in zero-voltage regime, when bias limiting JJs are not switching, and the least energy-efficient operation, when the bias limiting JJs are switching, providing the circuit with protection from the excessive bias current [36]. In the latter regime, when non-zero voltage is observed on the bias limiting junctions, the circuit still functions correctly, albeit by dissipating some additional power in the bias-limiting JJs. On decreasing temperature, the bias margins for all ERSFQ PCMs in Fig. 5, from simplest DC/SFQ-JTL-SFQ/DC to most complex NDRO SW, monotonically decreased from as high as ±22% for the former PCM at 4.2 K to 5% for the latter one at 100 mK. Since PCMs were designed and optimized for 4.2 K operation, the change in margins can be also explained by the ~15% increase in $I_c$ of JJs from 4.2 K to 100 mK.

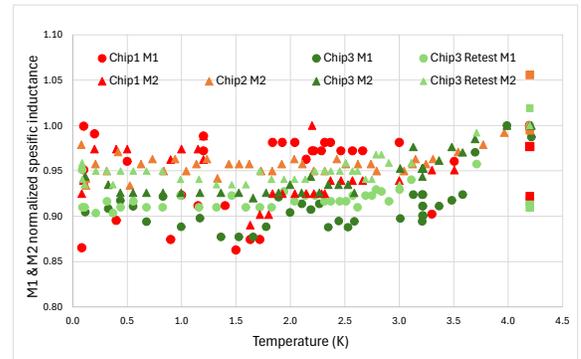

**Fig. 3.** Temperature dependence of normalized specific inductance for M1 and M2 Nb layers. Three different chips with these inductances were tested. See text for details.

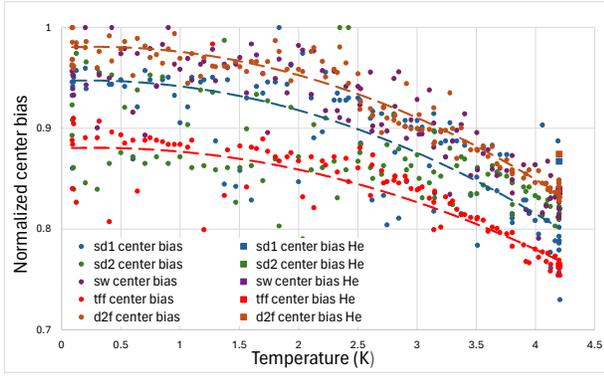

**Fig. 4.** Temperature dependence of normalized center bias for five tested ERSFQ digital PCMs. See text for details.

Figs. 6 and 7 show experimental details for testing the 1-to-4 demultiplexer and 8-bit programmable counter to be used in ERSFQ based digital qubit control [9, 17] (both NDRO based). Section IIA lists design details for the circuits and JJ counts. Fig. 6(b) and 7(b) display input and output signals of the functionality testing performed at ~2 MHz, with inputs supplied via current pulses sent to the chips' DC-to-SFQ input converters, and digitized outputs representing SFQ pulses as jumps in voltage level.

The DMX is programmed by a serial interface to a 4-cell memory register that corresponds to the 4 possible output channels. A '1' bit in a memory cell allows SFQ pulses generated by a shared input to route through the associated output channel. The test pattern used here is a 4-channel rotation pattern. The register is first cleared with 4 serial clock pulses, then a single selection data bit is latched into the first memory cell. Four pulses are sent to the input to confirm they pass through to the first output channel (and no other outputs). Then a serial clock pulse moves the '1' data bit to the next cell and the clock test is repeated for the second output. This process repeats until every cell of the DMX has been tested. The DMX test is considered a pass if no pulses are missing and there are also no extraneous pulses observed on a deselected channel. For the mK test, this full test pattern worked only intermittently with recurring problems observed on output 1. We believe this was caused by either trapped flux in the output 1 output monitor or a fabrication defect. The latter may also cause a nonzero error rate in the 4.2 K test and seems to be more plausible. Therefore, a simpler pattern was then tested where the register was cleared and only a single DMX channel was selected at a time with the serial interface, resulting in ±6% bias margins with zero error rate for outputs 2 and 3.

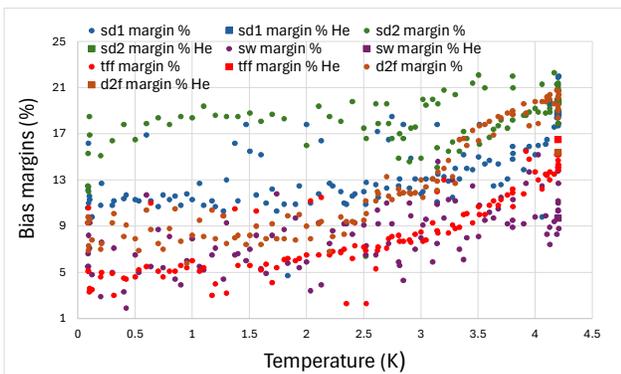

**Fig. 5.** Temperature dependence of operational total bias margins for five tested ERSFQ digital PCMs. See text for details.

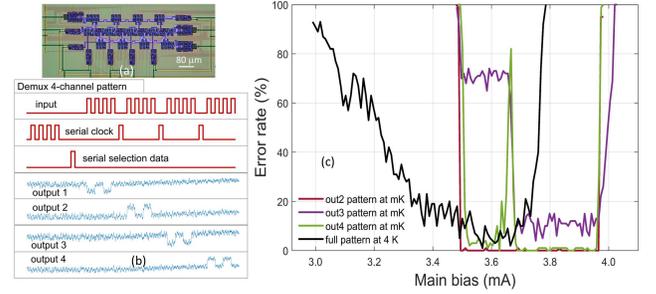

**Fig. 6.** Layout and results for 1-to-4 NDRO based demultiplexer: (a) circuit layout, (b) snapshot of input and output signals and (c) error rate vs bias current dependence.

For the PC, the number of pulses to be counted is programmed via a serial numeric data interface. Rising edges of the "serial data" signal latch a "1" bit into the first NDRO cell of the PC's shift register. A subsequent rising edge of the serial clock moves these serial data bits from one cell to the next. There are 8 cells total in the serial memory register, enabling counts of any number between 1 and 256. Although multiple patterns were tested, the test pattern presented in Fig. 7 programs the PC to count to 8, which fills the serial memory register with a mixture of 0 and 1 bits. This is a representative pattern because it will produce clear error signatures if there are either additional or missing pulses.

With the increase in circuit complexity from the DMX to the 8-bit counter, the PC only exhibited zero error rate with $\pm5\%$ bias margins at 4.2 K but never worked at 10 mK. This can be explained by the increase in $I_c$ by $\sim15\%$ bringing circuit bias outside of the operational margins. To bring the $I_c$ down, we employed an annealing process for $\sim30$ minutes at 230°C in ambient air. This allows for relatively quick $I_c$ adjustments, short of circuit redesign or fabrication process retargeting. By tuning annealing process parameters while simultaneously testing analog and digital PCMs, we were able to reduce the $I_c$ of the programmable counter by $\sim15\%$. After annealing, the counter operated with zero error with the same $\pm5\%$ margins as unannealed at 4.2 K.

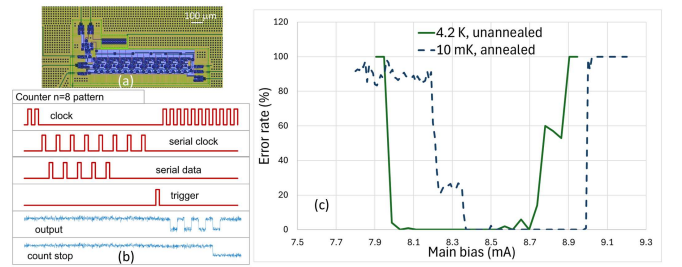

**Fig. 7.** Layout and results for 8-bit programmable counter: (a) circuit layout, (b) input and output signals and (c) error rate vs bias current dependence for n=8 pattern.

## IV. CONCLUSION

We report on systematic testing in the temperature range from 4.2 K to 10 mK of a comprehensive set of analog PCMs, ERSFQ cells, and more complex circuits such as demultiplexers and programmable counters used in digital qubit control. We compare the operating margins and error rates of these circuits and find that the center of bias margins (i.e., the optimal bias current value) for the tested circuits increases by $\sim15\%$ from 4.2 K to millikelvin. In the temperature range of interest, the measured Josephson junction (JJ) critical current $I_c$ of the PCM JJ arrays increases by $\sim15\%$, in good agreement with both theory and measured change of the center of bias margins for the tested digital circuits. In contrast, the geometrical and kinetic inductances (Nb and NbN layers, respectively) were not affected by the change in temperature. The operational bias margins for ERSFQ PCM and qubit control circuits designed to operate at 4.2 K decreased as temperature dropped, with this decrease in functional range getting more pronounced with an increase in circuit complexity. The operational margins at millikelvin completely disappeared for the most complex circuit, the programmable counter, and the circuit was not functional at any point in the measured bias range. With $I_c$ of the junctions of the counter reduced by $\sim15\%$ by means of annealing, the counter maintained operational bias margins at millikelvin comparable to the margins of the unannealed counter at 4.2 K.

Our results confirm that with simple adjustment, the ERSFQ circuits used for qubit control are able to operate at millikelvin temperatures with similar bias margins as at their design temperature of 4.2 K. This can be achieved by design modification to account for critical current change, retargeting of fabrication process critical current density, or post-processing annealing. Our results pave the way for high-speed and low-power superconducting electronics to be proximally integrated with qubits for digital control, readout, and co-processing for low-latency real-time QEC.

## REFERENCES

[1] M. Beverland, P. Murali, M. Troyer, et al., "Assessing requirements to scale to practical quantum advantage," arXiv:2211.07629, 2022.

[2] C. Gidney and M. Ekera, "How to factor 2048 bit RSA integers in 8 hours using 20 million noisy qubits," *Quantum*, vol.5, 433, 2021.

[3] S. Krinner, S. Storz, P. Kurpiers, et al., "Engineering cryogenic setups for 100-qubit scale superconducting circuit systems," *EPJ Quantum Technology* 6, 2 (2019).

[4] F. Battistel, C. Chamberland, K. Johar et al., "Real-time decoding for fault-tolerant quantum computing: progress, challenges and outlook," *Nano Futures*, vol. 7, 032003, 2023.

[5] D. Camps, E. Rrapaj, K. Klymko, et al., "Evaluation of the classical hardware requirements for large-scale quantum computations," in ISC High Performance 2024 Research Paper Proceedings (39th International Conference) (Prometeus GmbH, 2024) pp. 1–12.

[6] E. Charbon, M. Babaie, A. Vladimirescu and F. Sebastiano, "Cryogenic CMOS Circuits and Systems: Challenges and Opportunities in Designing the Electronic Interface for Quantum Processors," *IEEE Microwave Magazine*, vol. 22, no. 1, pp. 60-78, Jan. 2021.

[7] J. Yoo, Z. Chen, F. Arute et al., "Design and Characterization of a <4-mW/Qubit 28-nm Cryo-CMOS Integrated Circuit for Full Control of a Superconducting Quantum Processor Unit Cell," *IEEE Journal of Solid-State Circuits*, vol. 58, no. 11, pp. 3044-3059, Nov. 2023.

[8] R. McDermott, M. G. Vavilov, B. L. T. Plourde, et al., "Quantum-classical interface based on single flux quantum digital logic," *Quantum Sci. Technol.*, vol. 3, no. 2, 024004, 2018.

[9] O. Mukhanov, A. Kirichenko, C. Howington, et al., "Scalable Quantum Computing Infrastructure Based on Superconducting Electronics," 2019 *IEEE International Electron Devices Meeting (IEDM)*, San Francisco, CA, USA, 2019, pp. 31.2.1-31.2.4. doi: 10.1109/IEDM19573.2019.8993634.

[10] B. Weng, W. Peng, and J. Ren, "Low power single flux quantum qubit control circuit without high-frequency input," *Supercond. Sci. Technol.*, vol. 36, 095002, 2023.

[11] J. Barbosa, J. C. Brennan, A. Casaburi et al, "RSFQ All-Digital Programmable Multi-Tone Generator for Quantum Applications," *IEEE Trans. Quantum Engineering*, vol.6, 5500211, Dec. 2024.

[12] N. Takeuchi, T. Yamae, T. Yamashita, et al., "Microwave-multiplexed qubit controller using adiabatic superconductor logic," *npj Quantum Inf .*, vol. 10, 53, 2024.

[13] S. Reddy, M. Schmelz, J. Kunert et al., "Design and characterization of adiabatic quantum flux parametron using sub-µm cross-type Josephson junction technology," *Supercond. Sci. Technol.*, vol.38, 045002, Mar. 2025.

[14] F. Li, D. Pham, Y. Takeshita et al., "Energy Efficient Half-Flux-Quantum Circuit Aiming at Milli-Kelvin Stage Operation," *Supercond. Sci. Technol.*, vol.36, 105006, 2023.

[15] L. Di Palma, A. Miano, P. Mastrovito et al., "Discriminating the Phase of a Coherent Tone with a Flux-Switchable Superconducting Circuit," *Phys. Rev. Applied*, vol. 19, 064025, Jun. 2023.

[16] L. Di Marino, L. Di Palma, M. Riccio et al., "Control of a Josephson Digital Phase Detector via an SFQ-Based Flux Bias Driver," *IEEE Trans. Quantum Engineering*, vol.6, 3101708, 2025.

[17] J. Bernhardt, C. Jordan, J. Rahamim et al., "Quantum Computer Controlled by Superconducting Digital Electronics at Millikelvin Temperature," arXiv:2503.09879, Mar. 2025.

[18] E. Leonard, Jr., M. A. Beck, J. Nelson et al., "Digital Coherent Control of a Superconducting Qubit," *Phys. Rev. Applied* 11, 014009 (2019).

[19] C. Liu, A. Ballard, D. Olaya et al., "Single Flux Quantum-Based Digital Control of Superconducting Qubits in a Multichip Module, *PRX Quantum* 4, 030310 (2023).

[20] L. Di Palma, P. Mastrovito, A. Miano et al., "Fast Digital Phase Detection of a coherent tone at GHz frequencies," *IEEE Trans. Appl. Supercond.* vol. 34, No. 3, 2024, Art no. 1300305.

[21] D. Kirichenko, S. Sarwana, A. Kirichenko, "Zero Static Power Dissipation Biasing of RSFQ Circuits," *IEEE Trans. Appl. Supercond.*, vol. 21, no. 3, pp. 776-779, June 2011.

[22] A. Somoroff, P. Truitt, A. Weis, et al., "Fluxonium qubits in a flip-chip package," *Phys. Rev. Appl.*, vol. 21, 2024; Art. no. 024015.

[23] D. Yohannes, M. Renzullo, J. Vivalda, et al., "High Density Fabrication Process for Single Flux Quantum Circuits," *Appl. Phys. Lett.*, vol. 122, 2023, Art. no. 212601.

[24] D. Yohannes, S. Sarwana, S Tolpygo et al., "Characterization of HYPRES' 4.5 kA/cm² & 8 kA/cm² Nb/AlOₓ/Nb Fabrication Processes," *IEEE Trans. Appl. Supercond.*, vol. 15, 90-93 (2005).

[25] K. K. Likharev and V. K. Semenov, "RSFQ Logic/Memory Family: A New Josephson-junction Technology for Sub-Terahertz-Clock-Frequency Digital Systems," *IEEE Trans. Appl. Supercond.* 1, 3–28 (1991).

[26] S. V. Polonsky, V. K. Semenov, P. Bunyk et al., "New RSFQ Circuits," *IEEE Trans. Appl. Supercond.* 3, 2566-2577 (1993).

[27] S. V. Polonsky, V. K. Semenov, A. F. Kirichenko, "Single Flux Quantum B Flip-Flop and Its Possible Applications," *IEEE Trans. Appl. Supercond.* 4, 9-16 (1994).

[28] O. A. Mukhanov, S. V. Rylov, V. K. Semenov, and S. V. Vyshenskii, "RSFQ logic arithmetic," *IEEE Trans. Magnetics*, 25, 857-860 (1989).

[29] S. B. Kaplan and O. A. Mukhanov, "Operation of a superconductive demultiplexer using rapid single flux quantum (RSFQ) technology," *IEEE Trans. Appl. Supercond.*, vol. 5, 2853-2856 (1995).

[30] A. F. Kirichenko, A. Jafari-Salim, P. Truitt, et al., "System and method of flux bias for superconducting quantum circuits," U.S. Patent 20220399145A1, 2022.

[31] V. Ambegaokar and A. Baratoff, "Tunneling Between Superconductors," *Phys. Rev. Lett.*, vol. 10, 486, 1963.

[32] M. Tinkham, in *Introduction to Superconductivity*, (1996). Dover Publications, ch. 3, sec. 3.6.2, p. 63.

[33] S. K. Tolpygo, E. B. Golden, T. J. Weir et al., "Self-and mutual inductance of NbN and bilayer NbN/Nb inductors in a planarized fabrication process with Nb ground planes," *IEEE Trans. Appl. Supercond.* vol. 33, Feb. 2023, Art. no. 1101911.

[34] A. J. Annunziata, D. F. Santavicca, L. Frunzio, et al., "Tunable superconducting nanoinductors," *Nanotechnology* vol. 21, 445202 (2010).

[35] P. Foshat, S. Poorgholam-khanjari, V. Seferai, et al., "Quasiparticle dynamics in NbN superconducting Microwave resonators at single photon regime," *IEEE Trans. Quantum Eng.*, vol. 6, 2025, Art no. 3102608.

[36] C. Shawawreh, D. Amparo, J. Ren, et al., "Effects of Adaptive dc Biasing on Operational Margins in ERSFQ Circuits," *IEEE Trans. Appl. Supercon.*, 27(4), 1301606, June 2017.